\documentclass{article}
\usepackage{spconf,amsmath,graphicx}
\usepackage{multirow}
\usepackage{svg}
\usepackage{bbm}
\usepackage{float}

\title{MBNet: MOS Prediction for Synthesized Speech with Mean-Bias Network}

\name{Yichong Leng$^1$, Xu Tan$^2$, Sheng Zhao$^3$, Frank Soong$^2$, Xiang-Yang Li$^1$ and Tao Qin$^2$}
\address{$^1$University of Science and Technology of China,$^2$Microsoft Research Asia,$^3$Microsoft Azure Speech\\
$^1$lyc123go@mail.ustc.edu.cn,xiangyangli@ustc.edu.cn,\\
$^2$\{xuta,frankkps,taoqin\}@microsoft.com,$^3$sheng.zhao@microsoft.com}
\begin{document}
\ninept
\maketitle
\begin{abstract}
Mean opinion score (MOS) is a popular subjective metric to assess the quality of synthesized speech, and usually involves multiple human judges to evaluate each speech utterance. To reduce the labor cost in MOS test, multiple methods have been proposed to automatically predict MOS scores. To our knowledge, for a speech utterance, all previous works only used the average of multiple scores from different judges as the training target and discarded the score of each individual judge, which did not well exploit the precious MOS training data. In this paper, we propose MBNet, a MOS predictor with a mean subnet and a bias subnet to better utilize every judge score in MOS datasets, where the mean subnet is used to predict the mean score of each utterance similar to that in previous works, and the bias subnet to predict the bias score (the difference between the mean score and each individual judge score) and capture the personal preference of individual judges. Experiments show that compared with MOSNet baseline that only leverages mean score for training, MBNet improves the system-level spearmans rank correlation co-efficient (SRCC) by 2.9\% on VCC 2018 dataset and 6.7\% on VCC 2016 dataset.
\end{abstract}
\section{Introduction}
\label{sec:intro}

Speech quality assessment \cite{sqa} aims to measure the quality of synthesized speech and has long been a challenge in speech synthesis areas such as text to speech~\cite{tacotron2,fastspeech,ren2021fastspeech} and voice conversion~\cite{cycleganvc,autovc}. Existing objective metrics \cite{MCD,PESQ,ANIQUE,P563} can only reflect speech quality from some certain aspects. This motivates people to use subjective metrics, for instance MOS (mean opinion score) \cite{ana_vcc2016}, to give an overall human perception score. In a MOS test, multiple judges rate a discrete score between 1 and 5 for a given speech (the higher score, the better speech quality). The arithmetic mean of all judge scores of an utterance is the \textit{utterance-level MOS}. The mean of all utterance-level MOS of a speech-synthesis system is called \textit{system-level MOS}.

MOS test is expensive and time-consuming due to the involvement of human judges. To solve this problem, many methods have been proposed to automatically predict the MOS of an utterance: AutoMOS \cite{AutoMOS} predicted MOS with a long short term memory (LSTM) network. Quality-Net \cite{QualityNet} utilized frame-level quality constraint to stabilize the training procedure. With comparative experiments on several architectures, MOSNet \cite{mosnet} showed that CNN-BLSTM was a better architecture for MOS prediction. Based on MOSNet, Williams et al. \cite{williams2020} compared different speech representations for MOS prediction. 
Choi et al. \cite{choi2020neural} incorporated MOSNet with multi-task learning to improve performance. In \cite{choi2020deep}, global quality token and encoding layer were utilized to achieve better prediction accuracy.

Although many methods have been proposed to train the MOS prediction model and improve the prediction accuracy, there exists two important problems when dealing with the MOS training data:
\begin{itemize}
\item To the best of our knowledge, all the previous works ~\cite{QualityNet,mosnet,choi2020neural,choi2020deep} leveraged the utterance-level MOS as the training target, and discarded the detailed judge scores. Utterance-level MOS is simply an average score of multiple judge scores.
If we can leverage the individual score of each judge, the training data will be several times (depending on the number of judges on each speech) larger, which will be a large benefit due to the scarce of original training data. However, it is challenging to leverage these scores since human judgements are of high variance, which makes the model training unstable.
\item Utterance-level MOS is biased due to personal preference. In a MOS dataset, we always allocate more judges than the required number of judgements for each utterance\footnote{For instance, each audio in VCC 2018 MOS dataset is only rated by 4 judges but totally there are 267 judges since the huge number of participant systems.}, which can reduce the labor of each judge to ensure the judgement quality. However, different judges might have different rating standards due to the lack of quantitative criterion. As a result, different utterances may be rated by different sets of judges and thus receive incomparable utterance-level MOS. Directly training with the MOS data without any debiasing process, as adopted by many previous works, will hurt the model performance.
\end{itemize}

To address the above challenges, in this work, we propose MBNet, a MOS prediction model with a mean subnet and a bias subnet structure to leverage the individual judge scores of each utterance and handle the biased scores due to personal preference. Specifically, we design a mean subnet to predict the mean score similar to that in previous works, and a bias subnet to predict the bias score (the difference between the mean score and each judge score). Bias subnet can not only leverage all individual judge scores in dataset, but also capture the personal preference of each judge. The output of bias subnet is added with that of mean subnet to form the final prediction of individual judge score, which can force bias subnet to learn the bias of each judge only and thus stabilize the training. We further proposed clipped mean squared error loss to prevent model from over-fitting to the exact label, which might be incomparable. Experiments on VCC 2018 and VCC 2016 datasets show that MBNet significantly improves the prediction accuracy compared with previous MOS prediction systems including MOSNet and its variants \cite{mosnet,choi2020neural,choi2020deep}.

\section{Mean-Bias network}
\label{sec:methods}
In this section, we introduce the proposed mean-bias network (MBNet). We first describe the preliminary, and then introduce the framework as well as the specific designs of MBNet.

\subsection{Preliminary}
\label{ssec:prelimi}

Assuming a MOS dataset $D$ has $N$ audios. For each audio $a_i$ in $D$, there are $m$ \textbf{judge scores} $\{s_i^1, s_i^2, ...,s_i^m\}$ rated from $m$ judges, whose mean value is the utterance-level MOS $\Bar{s_i}$ (which is also called \textbf{mean score}). Note that there are totally $M$ ($M>m$) judges for the MOS test, and each utterance has $m$ judges which is a random subset of $M$ judges. Denote the delta between judge score $s_i^k$ and the mean score $\Bar{s_i}$ as \textbf{bias score} of judge $J_k$. 

Previous works usually regard the inconsistency of judge scores for an utterance as noise and use $(a_i, \Bar{s_i})$ as input-label pair to train the MOS prediction model ~\cite{mosnet,choi2020neural,choi2020deep}. The unused judge scores are quite a waste of expensive MOS dataset.

To reduce the labor of each judge, the total judges involved in a MOS test is usually much larger than the number of judges rating for each audio, i.e., $M \gg m$. In this way, two audios may be rated by totally different judges, leading to incomparable \textit{utterance-level MOS}
\footnote{Given that \textit{system-level MOS} is the mean of all \textit{utterance-level MOS} of the same speech-synthesizer, the \textit{system-level MOS} is comparable for practical usage because the number of judges rating for a system is large.}. The direct use of incomparable utterance-level MOS without any processing in many previous works will hurt the model training.

\subsection{MBNet Framework}
\label{ssec:mbnet}

\begin{figure}
  \centering
  \includegraphics[width=.42\textwidth]{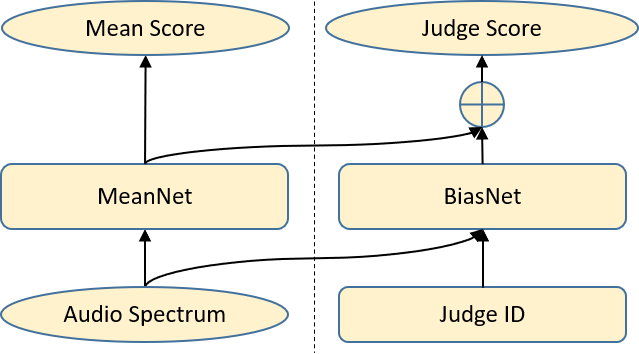}
  \caption{ \small An overview of the MBNet architecture consisting of a MeanNet (left part) and a BiasNet (right part). MeanNet predicts the mean score based on the audio spectrum. BiasNet predicts the bias score rated by a certain judge, conditioning on spectrum and judge ID, where the bias score is the delta value between the mean score and each judge score. The output of BiasNet is added with the output of MeanNet to form the final prediction of each judge score.}
  \label{fig:MBNet}
\end{figure}

To address the incomparable utterance-level MOS problem and better utilize the MOS dataset, we propose MBNet as shown in Fig. \ref{fig:MBNet}, which consists of two subnets: MeanNet and BiasNet. In general, the MeanNet is trained to predict the mean score and the BiasNet is trained to predict the bias score.

The MeanNet takes the spectrum of audio $a_i$ as input and uses mean score $\Bar{s_i}$ as label. The MeanNet is optimized in regression manner to predict MOS, the mean of judge scores. We use BiasNet to capture the bias of each judge. Apart from the spectrum of audio $a_i$, BiasNet also takes judge ID $J_k$ as input and use judge score $s_i^k$ rated by judge $J_k$ as label. In order to force BiasNet to learn the bias score of the given judge, we add the output of BiasNet with the output of MeanNet. By joint-training MeanNet and BiasNet, the label of BiasNet is set to bias score implicitly.

The judge score is a discrete value drawn from $\{1,2,3,4,5\}$. As a result, if one judge rates two audios with \textit{the same} judge score, it implies that from the perspective of this judge, the two audios are of \textit{the similar} quality rather than \textit{the same} quality. Therefore, it is not accurate to optimize the model to exactly output the judge score.
In MBNet, we propose clipped mean squared error (MSE) loss to prevent model from over-fitting to the exact judge score:
\begin{equation}
    \mathcal{L}_{C}( y, \hat{y}) = \mathbbm{1} (|y - \hat{y}| > \tau)(y-\hat{y})^2
    \label{eq:cmse}
\end{equation}
Where $\mathbbm{1}(\cdot)$ is the indicator function whose output is $1$ when the condition is true otherwise $0$, $y$ denotes the model output and $\hat{y}$ denotes the label, i.e., mean score or judge score. $\tau$ is the threshold below which the loss is set to $0$. In this way, the model does not struggle to fit an exact score but only the value in a loose range, which can improve the generalization of the MOS prediction. The final loss function of MBNet is:
\begin{small}
\begin{equation}
    loss = \mathcal{L}_C(W_{m}(a_i), \Bar{s_i}) + \lambda \mathcal{L}_C(W_{b}(a_i, J_k) + W_{m}(a_i), s_i^k)
    \label{eq:final}
\end{equation}
\end{small}Where $\mathcal{L}_C$ is the clipped MSE loss. $W_{m}, W_{b}$ are the parameters of MeanNet and BiasNet. $\lambda$ is the hyper-parameter to balance the training of MeanNet and BiasNet. $\mathcal{L}_C$ is also utilized in MeanNet to prevent MeanNet from over-fitting to the mean score, which could be biased due to small judge number. The loose range of $\mathcal{L}_C$ in MeanNet where the loss is set to $0$ is regarded as bias region and handled by BiasNet through joint-training.

\begin{figure}
  \centering
  \includegraphics[width=.48\textwidth]{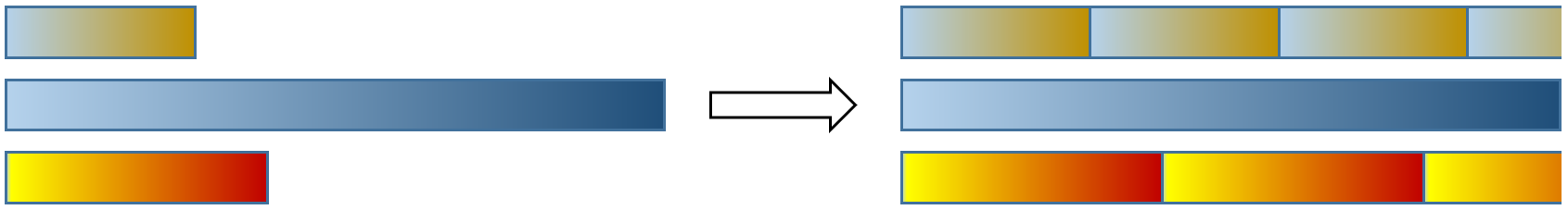}
  \caption{ Repetitive padding when formulating batch.}
  \label{fig:Reppad}
\end{figure}

Given that if we repeat an audio for multiple times, the quality, indicated by MOS, of the repeated audio should be the same as that of original audio. We propose repetitive padding in MBNet when formulating batch so as to stabilize the mean and variance estimation in batch normalization \cite{bn} compared with default zero-padding. As Fig. \ref{fig:Reppad} shows, in repetitive padding, the padding value is the replica of original speech.

It is worth noting that MBNet is a general framework to utilize judge scores in MOS dataset and capture the personal preference of judges, without any specific constraint on the detailed architectures of MeanNet and BiasNet.

\begin{figure}
  \centering
  \includegraphics[width=.45\textwidth]{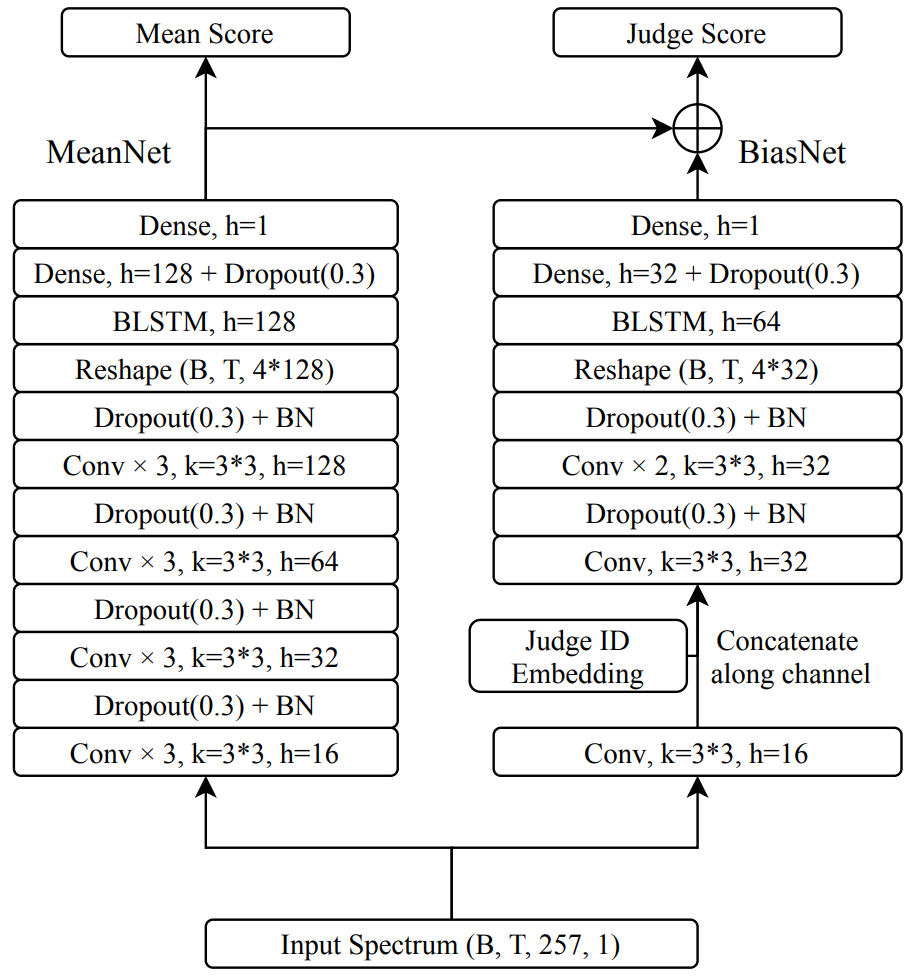}
  \caption{ \small The detailed architecture of MBNet used in experiments. We use 2D convolution layers (Conv), bidirectional LSTM (BLSTM) and batch normalization (BN) in our MBNet. In this figure, $k$ stands for kernel size, $h$ stands for the output hidden or channel size, $B$ stands for batch size and $T$ stands for the length of speech frames.}
  \label{fig:MBNet_detail}
\end{figure}

\section{EXPERIMENT SETTINGS}
\label{sec:exp}

In this section, we briefly describe the experiment settings and model details of MBNet.

\subsection{Dataset}
\label{ssec:dataset}
We use the MOS dataset from Voice Conversion Challenge 2018 (VCC 2018) \cite{vcc2018} to train MBNet. The VCC 2018 dataset contains 20580 audios submitted by 38 different systems. A total of 267 expert judges are involved in VCC 2018 and each audio is scored by 4 judges. We randomly split the dataset into training, validation and test sets with a size of 13580, 3000 and 4000. 

In order to test the generalization ability of MBNet, we further utilize the MOS dataset from Voice Conversion Challenge 2016 (VCC 2016) \cite{vcc2016} including 26028 audios from 20 systems as the test set. In order to match the large amount of audios in VCC 2016, we concatenate the training set and test set in VCC 2018 to form a new 17580-audio training set for MBNet, which is used for training when VCC 2016 serves as test set.

All audios are down-sampled to 16kHz and then converted to 257-dimension linear spectrum by short-time Fourier transform (STFT), and taken as the input to MBNet.

\subsection{Model Details of MBNet and Baseline System}
\label{ssec:mosnet}

The MeanNet consists of 4 convolution blocks. Each block contains 3 2D-convolution layers, followed by a dropout and batch normalization layer. For the last 2D-convolution layer of each block, the stride is set to (1, 3). A stack of 1 BLSTM layer and 2 dense layers is added after the convolution blocks. For BiasNet, the architecture is similar but smaller. Only 2 convolution blocks where each block contains 2 2D-convolution layers are utilized. The judge ID is mapped to embedding and concatenated with the input spectrum after the first convolution layer. The detailed architecture of MBNet is shown in Fig. \ref{fig:MBNet_detail}.

In both MeanNet and BiasNet, the output of dense layers is the score of each frame. The final score is the global mean pooling of frame-wise scores. MBNet utilizes the frame-level loss \cite{QualityNet} to force each frame to predict the right score, so as to stabilize the model output. The hyper-parameter $\tau$ in Equation \ref{eq:cmse} is set to $0.5$ to match the minimum granularity of judge score, i.e., $1.0$. We set $\lambda$ in Equation \ref{eq:final} to $4.0$ so as to let model focus more on BiasNet, which is relatively hard to train.

We choose MOSNet \cite{mosnet} with CNN-BLSTM architecture as our baseline system, which is the state-of-the-art MOS prediction framework. MOSNet directly uses the audio and its MOS in dataset to train a MOS predictor, without leveraging specific judge scores. The parameter amount of MBNet is a little bit more (12\%) than the MOSNet baseline. From our experiments\footnote{Not shown for the space limitation.}, directly adding 12\% parameter to MOSNet results in almost the same performance.

\begin{table*}[t]
\caption{The MOS prediction results of MBNet and MOSNet baseline on VCC 2018 and VCC 2016 dataset. The best results are highlighted in bold.}
\label{tab:ablation_result}
\begin{center}
\begin{tabular}{l|cccccc|ccc}
\hline
\multirow{3}{*}{Model} & \multicolumn{6}{c|}{VCC 2018}                                            & \multicolumn{3}{c}{VCC 2016}
\tabularnewline\cline{2-10}
\multicolumn{1}{c|}{}   & \multicolumn{3}{c|}{\textit{utterance-level}} & \multicolumn{3}{c|}{\textit{system-level}} & \multicolumn{3}{c}{\textit{system-level}} \\
\multicolumn{1}{c|}{}    &    LCC        & SRCC       & \multicolumn{1}{c|}{MSE}       & LCC       & SRCC      & MSE      & LCC       & SRCC       & MSE\\\hline\hline
MOSNet \cite{mosnet}      &   0.638  &  0.611   &  \multicolumn{1}{c|}{0.465}    &  0.964   &  0.922  &  0.047   &  0.915   &  0.862   &  0.308\\
MBNet &   \textbf{0.680}  &  \textbf{0.647}   &   \multicolumn{1}{c|}{\textbf{0.426}}    &  \textbf{0.977}   &  \textbf{0.949}  & \textbf{ 0.029}   &  \textbf{0.941}   &  \textbf{0.920}   &  \textbf{0.188} \\

\hline
\end{tabular}
\end{center}
\vspace{-0.6cm}
\end{table*}

\subsection{Training Configuration}
\label{ssec:model}
We train the MBNet with 1 NVIDIA Tesla V100 GPU. The batch size is set to 64. We use Adam optimizer \cite{adam} with $10^{-4}$ learning rate. We choose the checkpoint with the smallest validation loss among 50 epochs. We repeat all experiments for 10 times with different random seeds and report the averaged metrics including MSE, spearmans rank correlation coefficient (SRCC) \cite{SRCC} and linear correlation coefficient (LCC) \cite{LCC}. On VCC 2018 dataset, we report the utterance-level result and the system-level result. Due to the lack of utterance-level MOS in VCC 2016 dataset, only the system-level result is reported.

\section{RESULTS AND ANALYSES}
\label{sec:typestyle}

In this section, we first report the MOS prediction results of MBNet, and then give further analyses on our method. 

\subsection{MOS Prediction Results}
\label{ssec:training}
Table \ref{tab:ablation_result} is the MOS prediction results on VCC 2018 and VCC 2016 dataset. As the table shows, the proposed MBNet has a better result on both datasets than the MOSNet baseline. The improvement is consistent on LCC, SRCC and MSE. We use the official code to train the MOSNet.

In practice, the MOS is used to compare the quality of two speech-synthesis systems. Therefore, system-level SRCC, which is an indicator of relative rank of different systems, is the most important metric for MOS prediction. Our MBNet can improve system-level SRCC from 0.922 to 0.949 on VCC 2018 dataset. On VCC 2016 dataset, the improvement is more significant, i.e., from 0.862 to 0.920.

\subsection{Comparison with Other Methods}
Apart from the MOSNet, we compare MBNet with several strong MOSNet systems~\cite{choi2020neural,choi2020deep} incorporating other technologies such as multi-task learning (MTL), focal loss (FL) \cite{focalloss}, global quality token (GQT) and encoder layer (EL) \cite{encoding_layer}. Since the code of these systems are not released, we cannot know how they split the VCC 2018 dataset into training, validation and test set. Given that the different random split on VCC 2018 dataset could lead to different results, we only compare MBNet with these systems on VCC 2016 dataset \footnote{Models are trained on part of VCC 2018 dataset and test on the whole VCC 2016 dataset.}. In order to make sure MBNet cannot benefit from the split of dataset, we only use 13580 audios to train our MBNet compared with 15580 audios used in these systems. The results are shown in Table \ref{tab:mbnet_other}.

Note that the MBNet is trained with 17580 audios in Table \ref{tab:ablation_result} but 13580 audios in Table \ref{tab:mbnet_other}, which accounts for the performance difference. From the Table \ref{tab:mbnet_other}, MBNet can still surpass the performance of these improved versions of MOSNet even if MBNet is trained with 15\% data less, showing that the improvement of MBNet is quite significant. Although we compare MBNet with these systems, we want to emphasize that the technologies in these systems are not mutually exclusive with MBNet and can be utilized to further boost MBNet, which is an important future work of MBNet.

\subsection{Ablation Study}

We conduct ablation study to show the effectiveness of each component in MBNet. The results are shown in Table \ref{tab:mbnet_compare}. Due to space limitation, we only show SRCC results, which can better measure the MOS prediction performance for practical usage.

From the table, removing BiasNet from MBNet will hurt the performance, showing that BiasNet can boost the MeanNet by joint-training. Note the performance of the MeanNet alone (left part only in Figure \ref{fig:MBNet}) is higher than MOSNet baseline, suggesting that MeanNet has a better architecture. The performance of the bare BiasNet alone (right part only in Figure \ref{fig:MBNet}) is poor because the model over-fits to the judges, indicating that MeanNet is necessary because it can increase the generalization ability of BiasNet by forcing the BiasNet only learning the judge bias. The utterance-level SRCC of the BiasNet alone on VCC 2018 dataset is quite low, and we will analyze this phenomenon in the next subsection.

We also study the efficiency of proposed clipped MSE loss and repetitive padding. The results show that repetitive padding is better than the default zero-padding. Although clipped MSE loss could lead to performance loss in utterance-level result, it can improve the system-level result, which is more important in practice, by preventing model from over-fitting to the exact MOS score.

\begin{table}[t]
\caption{The comparisons between MBNet and other MOSNet systems with advanced technologies such as multi-task learning (MTL), focal loss (FL), global quality token (GQT) and encoder layer (EL). The results come from the original paper ~\cite{choi2020neural,choi2020deep}.}
\label{tab:mbnet_other}
\begin{center}
\begin{tabular}{l|l|ccc}
\hline
\multirow{2}{*}{Model} & \multicolumn{1}{c|}{Data}  & \multicolumn{3}{c}{VCC 2016}
\tabularnewline\cline{3-5}
\multicolumn{1}{c|}{} & \multicolumn{1}{c|}{Size}   & LCC       & SRCC       & MSE   \\\hline\hline

MOSNet ~\cite{choi2020neural,choi2020deep} & 15.6k    &   0.896  &  0.858   &  0.316    \\ 
MOSNet+MTL \cite{choi2020neural}& 15.6k    &  0.925  &  0.883   &  0.227    \\ 
MOSNet+MTL+FL \cite{choi2020neural}& 15.6k    &   0.904  &  0.864   &  0.208    \\
MOSNet+GQT  ~\cite{choi2020deep}& 15.6k    & 0.921  &  0.853   &  0.242    \\
MOSNet+EL  ~\cite{choi2020deep}& 15.6k    & 0.908  &  0.855   &  0.242    \\\hline\hline
MBNet  & 13.6k & \textbf{0.931}   &   \textbf{0.906}   &  \textbf{0.207}   \\

\hline
\end{tabular}
\end{center}
\vspace{-0.5cm}
% \end{small}
\end{table}

\begin{table}[t]
\caption{Ablation study of MBNet, where Reppad stands for repetitive padding and CMSE stands for clipped MSE loss. We only report the SRCC metric due to the space limitation.}
\label{tab:mbnet_compare}
\begin{center}
\begin{tabular}{l|ccc}
\hline
\multirow{2}{*}{Model}  & \multicolumn{3}{c}{SRCC}
\tabularnewline\cline{2-4}
\multicolumn{1}{c|}{}   & Utterance 18'       & System 18'       & System 16'   \\\hline\hline
MBNet       &   0.647  &  \textbf{0.949}   &  \textbf{0.920}    \\
$-$ BiasNet &   0.633  &  0.934   &  0.916    \\
$-$ MeanNet &   0.343  &  0.887   &  0.873    \\
$-$ CMSE    &   \textbf{0.652}  &  0.946   &  0.915    \\
$-$ Reppad  &    0.644       &  0.929     &  0.912         \\
\hline
\end{tabular}
\end{center}
\vspace{-0.5cm}
\end{table}

\subsection{Bias Learned in BiasNet}

The following experiment is completed to verify that the BiasNet indeed learns the bias of every judge.
During the inference phase of MBNet, we can predict MOS by MeanNet alone or MeanNet plus BiasNet (MeanNet+BiasNet). If BiasNet captures the bias of each judge, then MeanNet+BiasNet should have better MOS prediction performance with the correct judges (the 4 judges who actually rate the given audio in dataset) provided. We compare the utterance-level results on VCC 2018 dataset in 3 conditions: 1) Predict MOS by MeanNet, 2) Predict MOS by MeanNet+BiasNet with random judges, 3) Predict MOS by MeanNet+BiasNet with correct judges. The results are shown in Table \ref{tab:mbnet_judge}. 

\begin{table}[t]
\caption{The utterance-level results of MBNet on VCC 2018 dataset with different test conditions.}
\label{tab:mbnet_judge}
\begin{center}
\begin{tabular}{l|ccc}
\hline
\multirow{2}{*}{MBNet predicts MOS by}  & \multicolumn{3}{c}{VCC 2018}
\tabularnewline\cline{2-4}
\multicolumn{1}{c|}{}   & LCC       & SRCC       & MSE   \\\hline\hline
MeanNet                 &   0.680  &  0.647   &  0.426    \\
MeanNet+BiasNet + Random Judge&   0.560  &  0.527   &  0.568    \\
MeanNet+BiasNet + Correct Judge &   \textbf{0.753}  &  \textbf{0.740}   &  \textbf{0.339}    \\
\hline
\end{tabular}
\end{center}
\vspace{-0.6cm}
\end{table}

It can be seen that if the MeanNet+BiasNet is provided with the correct judges, the utterance-level SRCC performance improves largely from 0.647 to 0.740, suggesting that BiasNet successfully captures the judge bias. However, if random judges are provided, the SRCC drops from 0.647 to 0.527, indicating that the variance of judge opinion is indeed very significant, which also accounts for the low utterance-level SRCC of the BiasNet alone in Table \ref{tab:mbnet_compare}.

\section{Conclusion}
\label{sec:conclusion}
In this paper, we proposed a novel MBNet for MOS prediction, which consists of a MeanNet to predict the mean score of each utterance and a BiasNet to capture the delta between the mean score and each specific judgement score. MBNet can not only capture the personal preference of each judge, but also fully utilize the limited MOS dataset. The MBNet framework is general and has no restriction to the model architectures of MeanNet and BiasNet. Experiments on VCC dataset show the significant improvements of MBNet over previous MOS prediction systems. For future works, we will consider combining our method with a better MOS prediction backbone so as to yield better prediction results, and also consider the prediction of other MOS variations including CMOS and SMOS.

\bibliographystyle{IEEEbib}
\bibliography{strings,refs}

\end{document}